\documentclass[fleqn,11pt]{article}
\usepackage{latexsym,amsfonts,epsfig,color}
\usepackage{times}

\newlength{\dinwidth}
\newlength{\dinmargin}
\newcommand{\Ga}{\alpha}

\newcommand{\Gd}{\delta}

\newcommand{\Gl}{\lambda}

\newcommand{\GO}{\Omega}
\newcommand{\Go}{\omega}

\newcommand{\Gs}{\sigma}

\newcommand{\Gt}{\tau}

\newcommand{\Gth}{\theta}
\newcommand{\GtH}{\vartheta}

\newcommand{\Gw}{\omega}

\newcommand{\CO}{{\cal O}}

\newcommand{\CZ}{{\cal Z}}

\newcommand{\bbC}{{\mathbb{C}}}

\newcommand{\bbP}{{\mathbb{P}}}

\newcommand{\bbZ}{{\mathbb{Z}}}

\newcommand{\scT}{{\scriptscriptstyle T}}

\newcommand{\indA}{{\scriptstyle A}}
\newcommand{\indB}{{\scriptstyle B}}

\newcommand{\iA}{{\scriptscriptstyle A}}
\newcommand{\iB}{{\scriptscriptstyle B}}
\newcommand{\iC}{{\scriptscriptstyle C}}
\newcommand{\iD}{{\scriptscriptstyle D}}

\newcommand{\iT}{{\scriptscriptstyle T}}

\newcommand{\TF}{{\widetilde F}}
\newcommand{\TP}{{\widetilde P}}
\newcommand{\TZ}{{\widetilde Z}}

\newcommand{\HA}{{\widehat A}}
\newcommand{\HH}{{\widehat H}}
\newcommand{\HT}{{\widehat T}}

\newcommand{\ft}[2]{{\textstyle {\frac{#1}{#2}} }}

\newcommand{\dd}{\partial}
\newcommand{\tr}{{\rm tr \,}}
\newcommand{\diag}{{\rm diag \,}}
\newcommand{\res}{{\rm res \,}}
\newcommand{\End}{{\rm End \,}}
\newcommand{\ra}{\rightarrow}
\newcommand{\I}{{\rm i}}

\newcommand{\qed}{\hfill$\Box$\bigskip}

\newcommand{\be}{\begin{equation}}
\newcommand{\ee}{\end{equation}}
\newcommand{\ben}{\begin{displaymath}}
\newcommand{\een}{\end{displaymath}}
\newcommand{\ba}{\begin{eqnarray}}
\newcommand{\ea}{\end{eqnarray}}
\newcommand{\nn}{\nonumber}
\newcommand{\non}{\nonumber\\}

\newcommand{\mathon}{\mathversion{bold}}
\newcommand{\mathoff}{\mathversion{normal}}

\newcommand{\la}{\label}

\newcommand{\Ref}[1]{(\ref{#1})}

\makeatletter
\@addtoreset{equation}{section}
\makeatother

\newcommand{\vl}{{\vphantom{[}}}

\newcommand{\equ}{\!=\!}
\newcommand{\pls}{\!+\!}
\newcommand{\mis}{\!-\!}

\newcommand{\Vv}{{v}}
\newcommand{\Vvt}{{\tilde{v}}}
\newcommand{\Vw}{{w}}
\newcommand{\Vwt}{{\tilde{w}}}
\newcommand{\aaa}{{a}}

\newcommand{\Ra}{{p}}
\newcommand{\Rb}{{q}}

\newcommand{\betaN}{{\beta^\vl_{\scriptscriptstyle N}}}

\newtheorem{theorem}{Theorem}[section]

\begin{document}

\begin{flushright}
SPIN-2001/29\\
nlin.SI/0112016
\end{flushright}
\bigskip\medskip

\begin{center}
\mathon
{\bf \LARGE Schlesinger transformations 
and quantum $R$-matrices}
\vskip0.4cm
\mathoff

\bigskip
\bigskip

{\bf\large N.~Manojlovi\'c\,$^a$ and H.~Samtleben\,$^b$}
\bigskip

$^a$\'Area Departamental de Matem\'atica, F. C. T.,
Universidade do Algarve\\ Campus de Gambelas, 8000-117 Faro,
Portugal\\
{\small E-mail: nmanoj@ualg.pt }\\[6pt]
$^b$ Spinoza Insituut, Universiteit Utrecht, Leuvenlaan 4\\
Postbus 80.195, 3508 TD Utrecht, The Netherlands \\
{\small E-mail: H.Samtleben@phys.uu.nl }
\end{center} 
\vskip0.6cm
 
\begin{abstract}
Schlesinger transformations are discrete monodromy preserving symmetry
transformations of a meromorphic connection which shift by integers
the eigenvalues of its residues. We study Schlesinger transformations
for twisted $\mathfrak{sl}_N$-valued connections on the torus. A
universal construction is presented which gives the elementary
two-point transformations in terms of Belavin's elliptic quantum
$R$-matrix. In particular, the role of the quantum deformation
parameter is taken by the difference of the two poles whose residue
eigenvalues are shifted. Elementary one-point transformations (acting
on the residue eigenvalues at a single pole) are constructed in terms
of the classical elliptic $r$-matrix.

The action of these transformations on the $\tau$-function of the
system may completely be integrated and we obtain explicit expressions
in terms of the parameters of the connection. In the limit of a
rational $R$-matrix, our construction and the $\tau$-quotients reduce
to the classical results of Jimbo and Miwa in the complex plane.

\end{abstract} 

\bigskip

\begin{center}
\end{center}
\vskip0.6cm

\newpage

\section{Introduction}

In this article we pursue the analysis of monodromy preserving
deformations of ordinary matrix differential equations of the type
\be
\frac{d\Psi(\Gl)}{d\Gl}= A(\Gl)\,\Psi(\Gl)
\;, 
\la{A} 
\ee 
for meromorphic connections $A(\Gl)$ on the torus. Isomonodromic
deformations of the system \Ref{A} in the complex plane have a long
history, dating back to the classical work of Schlesinger
\cite{Schl12}. They have been extensively studied in the early
eighties by Jimbo, Miwa and their collaborators
\cite{JiMiMoSa80,JiMiUe81,JimMiw81b,JimMiw81c}. Generalization to
higher genus Riemann surfaces has been adressed e.g.\
in~\cite{Okam86,Iwas92,Kawa97,KorSam97,Taka98,LevOls99}. The continous
isomonodromic deformations of \Ref{A} give rise to the system of
Schlesinger equations for the parameters of the connection $A$ as
functions on the moduli space of the Riemann surface. A central object
associated with these deformation equations is the $\tau$-function ---
the generating function of the Hamiltonians governing the deformation
flows. It turns out to be closely related to the Fredholm determinant
of certain integral operators associated to the Riemann-Hilbert
problem, see e.g.~\cite{HarIts97,BorDei01} for recent developments.

In addition to these continous isomonodromic deformations, there exist
discrete monodromy preserving transformations of \Ref{A}, which shift
by integers the eigenvalues of the residues of~$A$. In particular,
they intertwine between different solutions of the Schlesinger
equations. These discrete so-called Schlesinger transformations act by
left multiplication on the $\Psi$-function
\be
\Psi(\Gl)~\ra~F(\Gl)\,\Psi(\Gl)
\;,
\la{R1}
\ee
and subsequent normalization to keep $\det\Psi=1$. For the complex
plane, these transformations were systematically studied in
\cite{JimMiw81b,JimMiw81c}. Being written in terms of the
$\tau$-functions, their superposition laws provide a big supply of
discrete integrable systems.

Together with Korotkin, we have started in~\cite{KoMaSa00} the study
of Schlesinger transformations for twisted $\mathfrak{sl}_2$-valued
connections on the torus; the corresponding isomonodromic system has
been derived by Takasaki in~\cite{Taka98}. With the present article we
generalize the results of~\cite{KoMaSa00} to $\mathfrak{sl}_N$ with
an arbitrary $N$, while it is our primary purpose to elucidate their
general structure. As a main result, we find a universal formulation
of the Schlesinger transformations in terms of the quantum $R$-matrix
of the underlying integrable system. With a rational $R$-matrix, this
construction reproduces the classical results of Jimbo and Miwa
\cite{JimMiw81b} in the complex plane; the problem on the torus is
solved by means of Belavin's elliptic quantum
$R$-matrix~\cite{Bela81}. Schematically, this correspondence may be
sketched as
\begin{figure}[h]
  \begin{center}
    \leavevmode
     \input{RS.pstex_t}
  \end{center}
\end{figure}

Let us describe this in more detail. We consider the elementary
two-point Schlesinger transformation which lowers the $p$-th
eigenvalue of the residue of $A(\Gl)$ at $\Gl\!=\!\Gl_k$ and raises
the $q$-th of its residue eigenvalues at $\Gl\!=\!\Gl_l$. Denote its
multiplier matrix \Ref{R1} by $F(\Gl)=F_{k,l}^{p,q}(\Gl)$. The
ensemble of multiplier matrices $F_{k,l}(\Gl)\equiv \{
F_{k,l}^{p,q}(\Gl) |\, p,q \in\underline{N} \}$ gives a tensor in
$\End(\bbC^N)\otimes\End(\bbC^N)$, its action on the second
(auxiliary) $\bbC^N$ being labeled by the columns $p, q$ in which the
eigenvalues are shifted. We find that for the isomonodromic
deformations on the torus, this tensor may be expressed in terms of
Belavin's elliptic quantum $R$-matrix as~\footnote{Here, we use
standard tensor notation, explicitly defined in equation~\Ref{tensor}
of the appendix.}
\be
\stackrel{12}{F}{}\!\!^\vl_{k,l}\stackrel{\vl}{(\Gl)} ~=~\;
\stackrel2{G}{}\!\!_k^{-1} \stackrel{12}{R_{\vl}}{}\!\!\!
\stackrel{\vl}{(\Gl\mis\Gl_k,\Gl_{kl})}
\;\stackrel2{G}\!\!_l^\vl \;.
\la{R2}
\ee
The auxiliary space on the r.h.s.\ is conjugated with the matrices
$G_{k,l}$ which diagonalize the corresponding residues of the
connection $A(\Gl)$ and appear as parameters in the local expansion of
$\Psi$, cf.~\Ref{asymp} below.  The quantum deformation parameter in
\Ref{R2} is given by the difference $\Gl_{kl}=\Gl_k\!-\!\Gl_l$ of the
two poles whose residue eigenvalues are shifted.

This gives an interesting link between two objects of historically
rather different origin: the discrete Schlesinger transformations
originating from \Ref{A}, and the quantum $R$-matrix encoding the
Boltzmann weights of the $\bbZ_N\times\bbZ_N$ symmetric generalization
of Baxter's eight vertex model \cite{Bela81,ChuChu81} 
\be
\stackrel{12}{R}{}\!\!\stackrel{\vl}{(\Gl,\zeta)}~=~ 
\frac{\Gth_{[00]}(\frac{\zeta}{N})}
{\Gth_{[00]}(\Gl\pls\frac{\zeta}{N})}
\sum_{(\iA,\iB)\,\in\, \bbZ_N\times\bbZ_N}\!\!\! 
W_{\iA\iB}(\Gl,\zeta)\,
\stackrel1{\Gs}{}\!\!_{\iA\iB}\stackrel2{\Gs}{}\!^{\iA\iB}
\;,
\la{RRI}
\ee
with elliptic functions $W_{\iA\iB}$ and matrices $\Gs_{\iA\iB}$
defined in Appendix~A below, cf.~\Ref{R}. Inserting instead in
\Ref{R2} the rational $\mathfrak{sl}_N$ quantum $R$-matrix
\be
\stackrel{12}{R}{}\!\!\stackrel{\vl}{(\Gl,\zeta)}~=~ 
\frac{1}{N\Gl+\zeta}\;\Big(
\stackrel{\vl}{\Gl}\,  \stackrel{12}{I} ~+~
\stackrel{\vl}{\zeta}\;  \stackrel{12}{\Pi} \,\Big)
\;,
\la{RRr}
\ee
gives back the result of Jimbo and Miwa in the complex plane
\cite{JimMiw81b}. The correspondence is the more surprising as only
few properties of the $R$-matrix --- apart from its twist properties
\Ref{Rtwist}, these are essentially its standard value at zero
\Ref{R0} and the existence of an antisymmetric singular value
\Ref{Rsym} --- are sufficient to prove \Ref{R2}. In particular, the
remaining properties of the $R$-matrix, such as unitarity, crossing
symmetry, and most notably the quantum Yang-Baxter equation, by means
of \Ref{R2} directly translate into relations between Schlesinger
transformations --- supplying the latter with some hidden additional
(braiding) structure which remains to be understood and explored. The
canonical form of \Ref{R2} moreover suggests a possible extension of
the construction to other integrable systems, in particular to the
isomonodromic deformations on higher genus Riemann surfaces, see
e.g.~\cite{LevOls99}.

As the difference of the two poles $\Gl_k$ and $\Gl_l$ enters \Ref{R2}
in the form of the quantum deformation parameter of the $R$-matrix,
one may expect, that the elementary one-point transformations (which
shift two of the eigenvalues of a single residue) allow a construction
in terms of the {\em classical} elliptic $r$-matrix. Indeed, we find
that they are described by
\be
\stackrel{12}{F}{}\!\!^\vl_{k,k}\stackrel{}{(\Gl)} ~:=~ 
- \stackrel{2}{Y}\!\!^\vl_k \;+\,
\stackrel2{G}{}\!\!_k^{-1} \stackrel{12}{r\vphantom{R}}{}\!\!\!
\stackrel{}{(\Gl\mis\Gl_k)}
\;\stackrel2{G}\!\!_k^\vl\;,
\la{R3}
\ee
where the matrix~$Y_k$ is extracted from the second order in the local
expansion of $\Psi$ around~$\Gl_k$, cf.~\Ref{asymp}, and the classical
$r$-matrix is obtained in the $\zeta\!\ra\!0$ limit of \Ref{RRI},
cf.~\Ref{zeta0}. This gives the bottom row in the above sketched
correspondence. The structure of \Ref{R3} shows some similarity with
Sklyanin's universal form of B\"acklund transformations. The latter
has served to prove canonicity of B\"acklund transformations and to
shed light on their quantum counterparts
\cite{Skly00,Skly99,Skly00b}. Likewise, the form of \Ref{R3} --- and
more general also that of the two-point transformations \Ref{R2} in
terms of the quantum $R$-matrix --- seems adequate to address these
topics in the Schlesinger context. Quantization of the Schlesinger
system leads to some version of the Knizhnik-Zamolodchikov-Bernard
(KZB) equations \cite{Resh92,Harn94,KorSam97,Taka98}.~\footnote{More
specifically, quantization of the elliptic Schlesinger system
\cite{Taka98} which we are going to consider, leads to Etingof's
elliptic version of the KZB equations \cite{Etin94} which naturally
arises in the twisted WZW model of \cite{KurTak97}.}  It is then
tempting to speculate about quantum Schlesinger transformations
intertwining between solutions of the KZB equations in different
representations.

The Schlesinger transformations \Ref{R2}, \Ref{R3} act as solution
generating transformations of the elliptic Schlesinger
equations. Since the entire information about this system is encoded
in its $\Gt$-function, it is a natural question to which extent the
action of the Schlesinger transformations on the $\Gt$-function can
explicitly be described. In the complex plane, this action may
completely be integrated in terms of the parameters of the connection
and explicit rational functions \cite{JimMiw81b}. As second main
result of this paper we obtain from \Ref{R2}, \Ref{R3} a universal
formula for the change of the $\tau$-function under Schlesinger
transformations on the torus. For the elementary two-point
transformation \Ref{R2} the ratio of the transformed and the old
$\tau$ function is given by
\be
\frac{\widehat{\tau}}{\tau} ~=~ 
\betaN\cdot\sqrt[N]{\Ga}\;,
\qquad\quad
\Ga:=
\ft{\dd}{\dd\Gl}\,[\det F_{k,l}^{p,q}(\Gl)]_{\Gl=\Gl_l} \;,
\la{R4}
\ee
where $\Ga$ contains the entire functional dependence of the
$\tau$-quotient on the parameters of the connection, and $\betaN$
denotes an explicit function of $\Gl_{kl}$ and the modulus of the
torus, which is given in \Ref{beta} below. Again, for a rational $R$
matrix this formula reproduces the result of \cite{JimMiw81b}. The
proof of \Ref{R4} mainly relies on the property
\Ref{Rrel} of the quantum $R$-matrix.
For the elementary one-point transformations \Ref{R3} we similarly
obtain a change of the $\tau$-function by
\be
\frac{\widehat{\tau}}{\tau}  ~=~ \sqrt[N]{\Ga}  \;,
\qquad\mbox{with}\quad
\Ga:=\det F_{k,k}^{p,q}(\Gl) \;.
\la{R5}
\ee

The plan of this paper is as follows. Our setting is the isomonodromic
system of equations on the torus proposed by Takasaki
\cite{Taka98}. They are derived from \Ref{A} for meromorphic
$\mathfrak{sl}_N$-valued connections $A(\Gl)$ with constant cyclic
twists along $a$- and $b$-cycle.~\footnote{A different isomonodromic
system on the torus has been derived in \cite{KorSam97} allowing these
twists to vary with respect to the deformation parameters. The
isomonodromic deformation equations for these connections contain
transcendental dependence on the dynamical variables, which makes it
difficult to analyze this system in a way analogous to the Schlesinger
system on the sphere.} In Section~2 we recall following Takasaki the
notion of isomonodromic deformations for twisted meromorphic
connections on the torus, the associated elliptic Schlesinger system,
its symplectic realization and the definition of the
$\tau$-function. Section~3 is devoted to the construction of the
elementary two-point Schlesinger transformations~\Ref{R2} in terms of
Belavin's elliptic $R$-matrix, and the computation of their action on
the $\tau$-function \Ref{R4}. Finally, the analogous expressions for
the elementary one-point transformations \Ref{R3}, \Ref{R5} are
derived in Section~4.

\newpage

\section{Isomonodromic deformations on the torus}

\subsection{Schlesinger system}

We consider a meromorphic $\mathfrak{sl}_{N,\bbC}$-valued connection
$A(\Gl)$ with simple poles in $\Gl_1$, \dots, $\Gl_P$, which is
twisted along $a$- and $b$-cycle of the torus according to
\be
A(\Gl\pls1) = g^{-1}A(\Gl)\,g
\;,\quad
A(\Gl\pls\mu) = h^{-1}A(\Gl)\,h \;,
\la{twist}
\ee
with cyclic twist matrices $g, h$ explicitly given in \Ref{gh}. The
latter satisfy the exchange relations $\Go gh= hg$, where
$\Go=e^{2\pi\I/N}$ denotes the $N$-th root of unity.  A convenient
parametrization of $A(\Gl)$ is given in terms of the matrices
$\Gs_{\iA\iB}~:=~ h^\iA g^\iB$ and the following combinations of
Jacobi's theta functions
\be
w_{\iA\iB}(\Gl):= 
\frac{\Gth_{[\iA\iB]}(\Gl)\,\Gth_{[00]}'(0)}
{\Gth_{[\iA\iB]}(0)\,\Gth_{[00]}(\Gl)} \;,\qquad
\mbox{for}\quad (\indA,\indB)\not=(0,0) \;,
\ee
which have a simple pole with unit residue at $\Gl\!=\!0$; see
Appendix~A for details and further properties. The twisted connection
$A(\Gl)$ may be then parametrized in terms of its residues
\be
A_j ~:= \sum_{(\iA,\iB)\,\not=(0,0)}
A_j^{\iA\iB} \,\Gs_{\iA\iB} ~:=~ 
\res|_{\Gl=\Gl_j} A(\Gl) ~
\in ~\mathfrak{sl}_{N,\bbC}\;,
\la{res}
\ee
as
\be
A(\Gl) ~=~
\sum_{j=1}^P\; \sum_{(\iA,\iB)\,\not=(0,0)}
A_j^{\iA\iB} \,\Gs_{\iA\iB} \,w_{\iA\iB}(\Gl\mis\Gl_j) 
{}~=:~
\sum_{j=1}^{P}\;
\stackrel1{\tr}{}\!\left[\stackrel1{A_j}\, 
\stackrel{01}{r}{}\!(\Gl\mis\Gl_j)
\right] 
\;.
\ee
The second equality defines the classical elliptic $r$-matrix \Ref{r}
which here serves as a kernel to reconcile the given residues with the
correct global twist behavior.

We will study isomonodromic deformations of the system \Ref{A}:
\be
\frac{d\Psi}{d\Gl}=A(\Gl)\,\Psi\;.
\la{psi}
\ee
As in the complex plane, the asymptotical expansion of the matrix
$\Psi(\Gl)$ near the singularities $\Gl_j$ is of the form
\be
\Psi(\Gl)~=~ G_j\,\Big(I+(\Gl\mis\Gl_j)\cdot Y_j(\Gl\mis\Gl_j)\Big) \,
(\Gl-\Gl_j)^{T_j}\, C_j\;,
\la{asymp}
\ee
with constant matrices $G_j, C_j, T_j$ of which the former two are
elements of $SL(N,\bbC)$, and the latter one is traceless diagonal
\be
T_j = \diag\Big\{t_j^{(1)}, \dots, t_j^{(N)}\Big\}  \;.
\la{T}
\ee 
In the sequel we shall consider the generic case when none of the
eigenvalues of $T_j$ is integer or half-integer.  The
$\mathfrak{gl}_{N,\bbC}$-valued function $Y_j(\Gl)$ is holomorphic at
$\Gl\!=\!0$. The connection $A(\Gl)$ has the local form
\be
A(\Gl) = 
\frac{G^{\phantom{1}}_j T^{\phantom{1}}_j G_j^{-1}}{\Gl-\Gl_j}
+ G^{\phantom{1}}_j
\Big( Y_j(0) + [Y_j(0),T_j] \Big) G_j^{-1}
+ \CO\left( (\Gl\!-\!\Gl_j)^2\right) \;,
\la{Alocal}
\ee
which in particular gives an expression for the residues \Ref{res} in
terms of the parameters of the local expansion \Ref{asymp}.  Upon
analytical continuation around $\Gl\!=\!\Gl_j$, the function
$\Psi(\lambda)$ changes by right multiplication with some monodromy
matrices $M_j$
\be
\Psi(\Gl) ~\ra~  \Psi(\Gl)\,M_j\;,\qquad
M_j~=~C_j^{-1}\, e^{2\pi \I T_j}\, C^{\phantom{1}}_j\;.
\la{Mj}
\ee
Moreover, $\Psi(\lambda)$ has monodromies around the $a$ and $b$ cycle
of the torus:
\be
\Psi(\Gl\pls1) ~=~  g^{-1}\Psi(\Gl)\,M_a\;,\qquad
\Psi(\Gl\pls\Gt) ~=~  h^{-1}\Psi(\Gl)\,M_b\;.
\la{Mab}
\ee
The assumption of independence of all monodromy matrices $M_j$, $M_a$,
$M_b$ of the positions of the singularities $\Gl_j$ and the modulus
$\mu$ of the torus is called the isomonodromy condition. It defines
the isomonodromic dependence of the residues $A_i$ on the parameters
$\Gl_j, \mu$ which gives the generalization of the classical
Schlesinger system \cite{Schl12} to twisted connections on the torus.

\begin{theorem} \cite{Taka98} The isomonodromy conditions
\be
\dd_i M_j ~=~ \dd_i M_a ~=~ \dd_i M_b ~=~ \dd_\mu M_j ~=~ 
\dd_\mu M_a ~=~ \dd_\mu M_b ~=~ 0\;,
\ee
induce the following dependence of the residues $A_i$ on the
parameters $\Gl_j, \mu$
\ba
\dd_j A_i &=& \Big[ A_i\,, 
\sum_{(\iA,\iB)\,\not=(0,0)} 
A_j^{\iA\iB}\,\Gs_{\iA\iB}\,w_{\iA\iB}(\Gl_{ij}) \,\Big]
\;,\qquad\mbox{for}\quad i\not=j
\;,
\non
\dd_i A_i &=& -\sum_{j\not=i} \Big[ A_i\,, 
\sum_{(\iA,\iB)\,\not=(0,0)} 
A_j^{\iA\iB}\,\Gs_{\iA\iB}\,w_{\iA\iB}(\Gl_{ij}) \,\Big]
\;,
\non
\dd_\mu A_i &=& -\sum_{j} \Big[ A_i\,, 
\sum_{(\iA,\iB)\,\not=(0,0)} 
A_j^{\iA\iB}\,\Gs_{\iA\iB}\,\CZ_{\iA\iB}(\Gl_{ij}) \,\Big]
\;,
\la{SchlesT}
\ea
to which we will refer as the elliptic Schlesinger system.

\end{theorem}

\paragraph{Proof:\,\,} 
The proof is obtained by straightforward calculation of the
compatibility conditions derived from combining \Ref{psi} with the
isomonodromic dependence of the $\Psi$-function
\be
\dd_i\Psi \; \Psi^{-1} ~=~ \sum_{(\iA,\iB)\,\not=(0,0)} 
A_i^{\iA\iB}\,\Gs_{\iA\iB}\,w_{\iA\iB}(\Gl\!-\!\Gl_{i}) 
\;,
\la{isoPsi}
\ee
and analogously for $\dd_\mu\Psi\, \Psi^{-1}$.
\qed

\subsection{Poisson structure}

It has been shown in~\cite{Taka98} that the elliptic Schlesinger
system~\Ref{SchlesT} admits a symplectic realization with respect to
Sklyanin's linear bracket
\be
\Big\{ \stackrel1{A}\!(\Gl) , \stackrel2{A}\!(\Gl') \,\Big\} ~=~ 
\Big[ \stackrel{12}{r}\!(\Gl\mis\Gl')\;, 
\stackrel1{A}\!(\Gl) \,+ \stackrel2{A}\!(\Gl') \,\Big] \;,
\la{Poisson}
\ee
where $r(\Gl)$ denotes the elliptic $r$-matrix \Ref{r},
satisfying the classical Yang-Baxter equation
\be
\Big[\stackrel{12}{r}\stackrel{\vl}{(\Gl-\Gl')}\;,\, 
\stackrel{13}r\stackrel{\vl}{(\Gl)}+ 
\stackrel{23}r\stackrel{\vl}{(\Gl')}\Big] 
+ \Big[\stackrel{13}r\stackrel{\vl}{(\Gl)}\;,\, 
\stackrel{23}r\stackrel{\vl}{(\Gl')}\Big] ~=~0 \;.
\la{CLYB}
\ee
For a meromorphic connection with simple poles, the symplectic
structure \Ref{Poisson} is equivalent to the ${\mathfrak{sl}}_{N}$
Kirillov-Kostant bracket on the residues \Ref{res}. The Hamiltonians
describing the deformation \Ref{SchlesT} with respect to the variables
$\Gl_i$ and to the module $\mu$ of the torus are given by the
following contour integrals of the current $\tr A^2(\Gl)$ around the
singularities and the $a$-cycle of the torus, respectively,
\cite{Taka98,KoMaSa00}:
\ba
H_i &=& \ft{1}{4\pi \I}\oint_{\Gl_i}\tr A^2(\Gl)d\Gl ~=~
\sum_{j\neq i} \sum_{(\iA,\iB)\,\not=(0,0)} 
A_j^{\iA\iB} A_i{}_{\iA\iB} \, 
w_{\iA\iB}(\Gl_{ij}) \;,
\non[1ex]
H_\mu &=& -\ft{1}{2\pi \I}\oint_a \tr A^2(\Gl)d\Gl ~=~
- \sum_{i,j} \sum_{(\iA,\iB)\,\not=(0,0)}
A_j^{\iA\iB} A_i{}_{\iA\iB} \, 
{\cal{Z}}_{\iA\iB}(\Gl_{ij}) 
\;,
\la{Hcon}
\ea
with $A_{j\,\iA\iB}:=N \Gw^{(\iA\iB)}\,A_j^{-\iA,-\iB}\,$,
cf.~\Ref{sigo}, and the functions $\CZ_{\iA\iB}$ defined in
\Ref{Z}. These Hamiltonians mutually Poisson commute which directly
follows from $\{\tr A^2(\Gl),\tr A^2(\Gl')\}=0$. The local behavior of
the current $\tr A^2(\Gl)$ around $\Gl_i$ is given by:
\be
\tr A^2(\Gl) = \frac{C_i}{(\Gl-\Gl_i)^2} + \frac{2\,H_i}{\Gl-\Gl_i}
+ \CO\left( (\Gl\!-\!\Gl_i)^0\right) \;,
\la{localAA}
\ee
with Casimirs
\be
C_i = \tr A_i^2 = \tr T_i^2 \;.
\la{Cas}
\ee

\begin{theorem}
The elliptic Schlesinger system \Ref{SchlesT} is a multi-time
Hamiltonian system with respect to the symplectic structure
\Ref{Poisson} and the Hamiltonians \Ref{Hcon}:
\be
\dd_iA_j ~=~ \{H_i,A_j\} \;,\qquad \dd_\mu A_j ~=~ \{H_\mu,A_j\}\;.
\la{dynH}
\ee
\end{theorem}
\qed

We define the $\Gt$-function associated with the elliptic Schlesinger
system \Ref{SchlesT} as generating function of the Hamiltonians
\Ref{Hcon} according to \cite{JiMiUe81}
\be
\dd_i \ln \Gt ~=~ H_i\;,\qquad \dd_\mu \ln \Gt ~=~ H_\mu \;.
\la{tau}
\ee
Consistency of this definition follows from the fact that the
Hamiltonian flows \Ref{dynH} commute.

\section{Two-point Schlesinger transformations}

The elementary two-point Schlesinger transformation in the complex
plane, which lowers by one unit the eigenvalue of the residue $A_k$ in
a column~$\Ra$ and likewise raises the eigenvalue of the residue $A_l$
in a column~$\Rb$, acts as
\be
T_k ~\ra~ \HT_k ~=~ T_k - P_\Ra  \;,\qquad
T_l ~\ra~ \HT_l ~=~ T_l + P_\Rb 
\;,
\la{Ttraf0}
\ee
on the matrices $T_j$ from \Ref{asymp}, \Ref{T}. The matrices $P_\Ra$,
$P_\Rb$ here denote diagonal projection matrices $(P_\Ra)_m{}^n
=\Gd_m^\Ra\,\Gd_\Ra^n$. On the torus, however, we have to deal with
$\mathfrak{sl}_{N,\bbC}$-valued connections rather than
$\mathfrak{gl}_{N,\bbC}$. This is due to the fact that the
isomonodromic dependence of the $\Psi$-function \Ref{isoPsi} is
described by a connection with single pole and twist according to
\Ref{twist}, which is necessarily traceless. 
In particular, the matrices $T_j$ need to remain traceless under
Schlesinger transformations. The proper two-point transformations on
the torus hence comprise a shift of the eigenvalues according to
\Ref{Ttraf0} and a subsequent projection onto the traceless part,
i.e.\ they act as
\be
T_k ~\ra~ \HT_k ~=~ T_k - P_\Ra  + \ft{1}{N}\,I \;,\qquad
T_l ~\ra~ \HT_l ~=~ T_l + P_\Rb  - \ft{1}{N}\,I
\;.
\la{Ttraf}
\ee
This transformation hence does not strictly preserve the monodromy
matrices but changes two of them by constant cyclic factors according
to $M_k\ra\Gw M_k$, $M_l\ra\Gw^{-1} M_l$.  As stated in the
introduction, the Schlesinger transformation inducing
\Ref{Ttraf} may be described by acting on $\Psi$ with a multiplier
matrix $F(\Gl)$ from the left and subsequent normalization.

\begin{theorem}
The elementary two-point Schlesinger transformation of the system
\Ref{SchlesT} which shifts the eigenvalues of the residues $A_k$, $A_l$
according to \Ref{Ttraf} is given by
\be
\Psi(\Gl)~\ra~ \widehat{\Psi}(\Gl)~=~ 
\frac{F(\Gl)}{\sqrt[N]{\det F(\Gl)}}\,\Psi(\Gl) \;,
\la{ST}
\ee
where the $GL(N,\bbC)$-valued multiplier matrix $F(\Gl)$ is defined
via Belavin's elliptic $R$-matrix \Ref{R}
\be
\stackrel{0}{F}{}\!\!\stackrel{}{(\Gl)} ~:=~
\stackrel{0}{F}{}\!_{k,l}^{p,q}\!\stackrel{}{(\Gl)} ~:=~
\stackrel1{\Vv}{}\!^\iT \stackrel{01}{R}{}\!\!
\stackrel{}{(\Gl\mis\Gl_k,\Gl_{kl})}
\,\stackrel1{\Vw} \;,
\la{F}
\ee
and the dependence of $F(\Gl)$ on the parameters of the
$\Psi$-function is completely contained in the vectors $\Vv$ and $\Vw$
which contract the auxiliary space on the r.h.s. They are defined as
functions of the parameters in the local expansions \Ref{asymp} around
$\Gl_k$ and $\Gl_l$
\be
(\Vv^\iT)^m ~:=~ (G_k^{-1} )_\Ra{}^m \;,\qquad
\Vw_m ~:=~ (G_l)_m{}^\Rb \;.
\la{vw}
\ee

\end{theorem}

\paragraph{Proof:} The proof consists of three parts: we need to show
that the transformation \Ref{ST} does not change the twist properties
\Ref{Mab} of $\Psi(\Gl)$, induces the proper change \Ref{Ttraf} in the
local expansions \Ref{asymp} around $\Gl_k$, $\Gl_l$, and finally does
not introduce additional singularities in $\Psi(\Gl)$.

Belavin's elliptic $R$-matrix is explicitly given in Appendix~A
\Ref{R} together with several of its properties. 
The proper twist behavior of $F(\Gl)$ follows directly from the twist
properties of the $R$-matrix \Ref{Rtwist}. The correct local behavior
around $\Gl_k$, $\Gl_l$ is deduced from its properties \Ref{R0}, and
\Ref{Rsym}, i.e.\ from its standard value at zero $R(0)=\Pi$, and the
existence of an antisymmetric singular point. Together with \Ref{vw}
this implies
\be
F(\Gl_k)\, G_k = F(\Gl_k)\, G_k\, P_\Ra \;,
\qquad
F(\Gl_l)\, G_l = 0 \;.
\la{projkl}
\ee
With the local expansion \Ref{asymp} this shows that multiplication of
$\Psi$ by $F$ induces a shift of $T_{k,l}$ according to
\ben
T_k ~\ra~ \HT_k ~=~ T_k + I - P_\Ra  \;,\qquad
T_l ~\ra~ \HT_l ~=~ T_l + P_\Rb  
\;,
\een
which after normalization of $\Psi$ by the determinant gives
\Ref{Ttraf}. It remains to ensure that \Ref{ST} does not introduce any
additional poles in $\Psi$. To this end consider $\det F(\Gl)$ which
is a single-valued function on the torus. Equations
\Ref{projkl} show that it has a zero of order $N\mis1$ at $\Gl_k$, a
simple zero at $\Gl_l$ and by construction (cf.\ \Ref{R}) its only
pole (of at most $N$-th order) at $\Gl_k-\ft1N\Gl_{kl}$. According to
Abel's theorem, this pole then is precisely of $N$-th order and $\det
F(\Gl)$ possesses no more zeros. The normalized Schlesinger
transformation \Ref{ST} hence does not induce any additional
singularities in~$\Psi$. This completes the proof.

For later use, we note that due to its zero and pole structure, $\det
F(\Gl)$ may be given explicitly as
\be
\det F(\Gl) ~=~ -\Ga\;
\frac{
\Gth_{[00]}(\ft{N-1}N\Gl_{kl})^N\,
\Gth_{[00]}(\Gl\!-\!\Gl_k)^{N-1}\,
\Gth_{[00]}(\Gl\!-\!\Gl_l)}
{\Gth_{[00]}(\Gl_{kl})^{N-1}\,
\Gth_{[00]}'(0)\,
\Gth_{[00]}(\Gl\!-\!\Gl_k\!+\!\ft1N \Gl_{kl})^N}\;,
\la{detF}
\ee
with a constant $\Ga$ which in particular contains the entire
functional dependence of $\det F(\Gl)$ on the vectors $\Vv$ and $\Vw$.
\qed

Acting as discrete monodromy preserving transformation on the $\Psi$
function, the Schlesinger transformation \Ref{ST} maps solutions of
the Schlesinger system \Ref{SchlesT} to new solutions
\be
A(\Gl) ~\ra ~ \widehat{A}(\Gl) ~\equiv~ 
\dd_\Gl \widehat{\Psi}(\Gl)\:\widehat{\Psi}(\Gl)^{-1} \;.
\la{AA}
\ee
As the entire information about a solution of the Schlesinger system
is contained in its $\Gt$-function, \Ref{AA} induces the
transformation
\be
\Gt ~\ra~ \widehat{\Gt} \;,
\ee
implicitly defined by integration of \Ref{tau}, in which the action of
the Schlesinger transformation is entirely encoded.  In the rest of
this section we will show that --- like in the complex plane
\cite{JimMiw81b} --- the change of the $\Gt$-function may explicitly
be integrated in terms of the parameters of the local expansion of
the $\Psi$-function \Ref{asymp}. We first state the result as
\begin{theorem}
Under the transformation \Ref{ST}, the $\Gt$-function as defined in
\Ref{tau} changes as
\be
\frac{\widehat{\tau}}{\tau} ~=~ 
\betaN\cdot\sqrt[N]{\Ga}\;,
\la{varT}
\ee
with 
\be
\Ga~=~\Ga(\Vv,\Vw,\Gl_{kl},\mu)~=~
\ft{\dd}{\dd\Gl}\,[\det F]_{\Gl=\Gl_l} \;,
\la{alpha0}
\ee
which in particular contains the entire functional
dependence of the $\Gt$-quotient on the parameters $\Vv$ and $\Vw$,
and the explicit function
\be
\betaN~=~ \betaN(\Gl_{kl},\mu) ~=~ 
\frac{\Gth_{[00]}((1\!-\!\ft1N)\Gl_{kl};\mu)\;
\Gth_{[00]}'(0;\mu)^{(N-2)/N}}
{\Gth_{[00]}(\ft1N\Gl_{kl};\mu)\;\Gth_{[00]}(\Gl_{kl};\mu)^{(N-2)/N}} 
\;.
\la{beta}
\ee
\end{theorem}

Before coming to the proof of this theorem, let us note that with the
rational $R$-matrix \Ref{RRr} inserted in \Ref{F}, \Ref{alpha0}, the
factor $\sqrt[N]{\Ga}$ reduces to the result of \cite{JimMiw81b} in
the complex plane, up to an explicit function in $\Gl_{kl}$ which is
due to the different normalization, discussed above. Similarly, by
brief calculation one may verify that \Ref{varT}, \Ref{alpha0} also
reproduces the elliptic $N=2$ case for which the result has been given
in \cite{KoMaSa00} explicitly in terms of elliptic functions. In
particular, it is $\beta^\vl_2=1$.

\paragraph{Proof:}

According to the definition of the $\Gt$-function \Ref{tau} the
statement of the theorem is equivalent to
\ben
\HH_i-H_i ~=~ \dd_i \ln \betaN + \ft1{N}\, \dd_i \ln \Ga \;,
\qquad
\HH_\mu-H_\mu ~=~ \dd_\mu \ln \betaN + \ft1{N}\, \dd_\mu \ln \Ga \;,
\een
where $\HH_i$, $\HH_\mu$ denote the Hamiltonians associated with the
transformed $\Psi$ function \Ref{ST}. According to \Ref{Hcon} they are
obtained from
\ba
\tr\HA^2(\Gl) &=&  \tr A^2(\Gl) +  
2\,\tr\!\left[\, \Xi_{\Gl}(\Gl)\,A(\Gl)\;\right]+ 
\tr\!\left[\, \Xi_{\Gl}^2(\Gl)\;\right] \;,
\nn
\ea
where $\Xi_{\Gl}(\Gl)$ is given by
\be
\Xi_{\Gl}(\Gl)~:=~
F^{-1}(\Gl)\, \dd_\Gl F(\Gl)- \frac1{N}\,
\tr\!\left[F^{-1}(\Gl)\, \dd_\Gl F(\Gl)\right] 
\;.
\la{XI}
\ee
Altogether we need to show that
\ba
\dd_i \ln \betaN + \ft1{N}\, \dd_i \ln \Ga 
&=& \res_{\Gl_i}
\Big(\,\tr\!\left[\, \Xi_{\Gl}(\Gl) A(\Gl)\;\right]+ 
\ft12\,\tr\!\left[\, \Xi_{\Gl}^2(\Gl)\,\right]\,\Big) \;,
\la{tshow1}
\\[1ex]
\dd_\mu \ln \betaN + \ft1{N}\, \dd_\mu \ln \Ga
&=& -\ft1{2\pi i}\,\oint_a d\Gl
\Big(\,2\, \tr\!\left[\, \Xi_{\Gl}(\Gl) A(\Gl)\;\right]+ 
\tr\!\left[\, \Xi_{\Gl}^2(\Gl)\,\right]\,\Big)
\;.
\la{tshow2}
\ea
We proceed in several steps. To compute the isomonodromic dependence
of $\Ga$ we find it convenient to first construct the Schlesinger
transformation $F^{-1}(\Gl)$ inverse to \Ref{ST}. We then prove
\Ref{tshow1} separately for the cases $i\not=(k,l)$, $i=k$, and $i=l$,
respectively, and finally show \Ref{tshow2}.

\paragraph{Inverse Schlesinger transformation:}

The inverse multiplier matrix $F^{-1}(\Gl)$ is proportional to the
multiplier matrix $\TF(\Gl)$ of the Schlesinger transformation which
reverses the transformation \Ref{Ttraf}. According to Theorem~3.1 the
latter is of the form
\be
\stackrel{0}{\TF}{}\!\!\!\stackrel{}{(\Gl)} ~=~
\stackrel1{\Vvt}{}\!^\iT \stackrel{10}{R}{}\!\!
\stackrel{}{(\Gl_l\mis\Gl,\Gl_{kl})}
\,\stackrel1{\Vwt} \;,
\la{FT}
\ee
where we have made use of \Ref{Ras} and the vectors $\Vwt$, $\Vvt$ now
depend on the parameters of the local expansions \Ref{asymp} of the
transformed function $\widehat\Psi$. Accordingly, it is
\be
F(\Gl)\,\TF(\Gl)~=~  \aaa(\Gl)\cdot I \;.
\la{FFT}
\ee
The function $\aaa(\Gl)$ which denotes the proportionality factor
between $F^{-1}(\Gl)$ and $\TF(\Gl)$ is single-valued on the torus
with simple zeros in $\Gl_k$, $\Gl_l$ and simple poles in
$\Gl_k-\ft1{N}\Gl_{kl}$, $\Gl_l+\ft1{N}\Gl_{kl}$. It is determined by
its pole structure it is determined up to a $\Gl$-independent
constant. We fix this constant by normalizing $\TF(\Gl)$ such that
\be
\aaa'(\Gl_l) ~=~ (\det F)'(\Gl_l) \;,
\ee
i.e. $\TF(\Gl_l)$ is the matrix of minors of $F(\Gl_l)$. Consider the
explicit expansion of \Ref{F} and \Ref{FT} around $\Gl=\Gl_l$
\ba
F(\Gl) &=& M + (\Gl\mis\Gl_l)\, X + \dots  \non
&\equiv& 
\stackrel1{\Vv}{}\!^\iT 
\stackrel{01}{R}{}\!\!
\stackrel{}{(-\Gl_{kl},\Gl_{kl})} \,\stackrel1{\Vwt} ~+~
(\Gl\!-\!\Gl_l)\, \stackrel1{\Vv}{}\!^\iT 
\stackrel{01}{R}{}\!'\!
\stackrel{}{(-\Gl_{kl},\Gl_{kl})}\,\stackrel1{\Vwt} ~+~ \dots 
\non[2ex]
\TF(\Gl) &=&  
P + (\Gl\mis\Gl_l)\, Z + \dots\non
&\equiv& 
\stackrel1{\Vvt}{}\!^\iT 
\stackrel{01}{\Pi}\;\stackrel1{\Vwt} ~-~ 
(\Gl\!-\!\Gl_l)\, \stackrel1{\Vvt}{}\!^\iT 
\stackrel{10}{R}{}\!'\!
\stackrel{}{(0,\Gl_{kl})} 
\,\stackrel1{\Vwt} ~+~ \dots 
\;.
\la{MXPZ}
\ea
Plugging this into equation \Ref{FFT} , one finds that $\Vwt\sim\Vw$
and may hence be normalized to $\Vwt=\Vw$. The vector $\Vvt$ is then
defined as polynomial function of $\Vw$ and $\Vv$~\footnote{More
specifically, the vector $\Vvt$ is a homogeneous polynomial of order
$N\!-\!1$ in $v$ and $N\!-\!2$ in $w$. It is only for $N\!=\!2$ that
$\Vvt\sim\Vv$ which considerably simplifies the calculation
\cite{KoMaSa00}. In the rational case \cite{JimMiw81b}, the
polynomial factors such that again $\Vvt\sim\Vv$.}
\be
\det \left[\stackrel1{\Vv}{}\!^\iT \stackrel{01}{R}{}\!\!
\stackrel{}{(-\Gl_{kl},\Gl_{kl})}
\,\stackrel1{\Vw} \right]_n^{m} ~=~
\Vw_n\,\Vvt^m ~=~
P_n{}^m \;.
\la{defP}
\ee
where by $[\dots]_m^n$ we denote the corresponding minor. Equation
\Ref{FFT} further implies
\ben
PM=MP=0\;,\qquad MZ+XP = \Ga I = ZM + PX\;,
\een
\be
\Ga~=~\ft{\dd}{\dd\Gl}\,[\det F]_{\Gl=\Gl_l}  ~=~
\tr[XP] ~=~
\stackrel{1}{\Vvt}{}\!^\iT 
\stackrel{2}{\Vv}{}\!^\iT \stackrel{12}{R}{}\!'\!
\stackrel{}{(-\Gl_{kl},\Gl_{kl})}
\,\stackrel1{\Vw}\,\stackrel{2}{\Vw} 
 \;.
\la{alpha}
\ee
Analogous relations may be obtained from the expansion around
$\Gl=\Gl_k$. In particular, it follows that $\ft1{\Ga}PX$ and
$\ft1{\Ga}ZM$ are projectors of rank 1 and $N\mis1$,
respectively. Moreover, \Ref{alpha} gives different equivalent
expressions for the factor $\Ga$ which we shall use to compute the
l.h.s.\ of \Ref{tshow1}, \Ref{tshow2}, i.e.\ the isomonodromic
dependence of $\Ga$. To this end, we note that according to
\Ref{defP}, the matrix $P=\TF(\Gl_l)$ depends on $M=F(\Gl_l)$ as
\ben
\Gd P ~=~ -\ft1{\Ga}\,\Big(
Z(\Gd M)P + P(\Gd M)Z - P\,\tr[Z\Gd M] - Z\,\tr[P\Gd M] \Big) \;.
\een
Further computation together with \Ref{alpha} then allows to express
the variation of $\Ga$ as
\be
\tr[X\Gd P] ~=~ \tr[Z\, \Gd M]
\quad \Longrightarrow\quad
\Gd\, \Ga ~=~ \tr[P\Gd X] + \tr[Z\, \Gd M]
\;.
\la{XdP}
\ee
Since $X$ and $M$ are explicit functions of the vectors $\Vv$ and
$\Vw$, formula \Ref{XdP} in particular gives the isomonodromic
dependence of $\Ga$ as function of the corresponding dependence of
$\Vv$ and $\Vw$. According to \Ref{Alocal}, \Ref{vw}, the latter
follows from \Ref{SchlesT} to be
\ba
\dd_j \Vw &=& 
-\sum_{(\iA,\iB)\,\not=(0,0)} 
A_j^{\iA\iB}\,\Gs_{\iA\iB}\,w_{\iA\iB}(\Gl_{lj}) \; \Vw \;,
\qquad j\not=l\;,
\non
\dd_l \Vw &=& 
\sum_{j\not=l} \sum_{(\iA,\iB)\,\not=(0,0)} 
A_j^{\iA\iB}\,\Gs_{\iA\iB}\,w_{\iA\iB}(\Gl_{lj}) \; \Vw \;,
\non
\dd_j \Vv^\scT  &=& 
\Vv^\scT \sum_{(\iA,\iB)\,\not=(0,0)} 
A_j^{\iA\iB}\,\Gs_{\iA\iB}\,w_{\iA\iB}(\Gl_{kj}) \;,
\qquad j\not=k\;,
\non
\dd_k \Vv^\scT  &=& 
-\Vv^\scT \;\sum_{j\not=k}\sum_{(\iA,\iB)\,\not=(0,0)} 
A_j^{\iA\iB}\,\Gs_{\iA\iB}\,w_{\iA\iB}(\Gl_{kj}) \;.
\la{implicit}
\ea
Similar expressions are obtained from \Ref{SchlesT} for the $\mu$
dependence of these parameters. Together with \Ref{XdP} these
relations allow to compute the $\dd\Ga$ terms on the l.h.s. of
\Ref{tshow1},\Ref{tshow2}. We are now in position to prove \Ref{tshow1}
for $i\not=(k,l)$, $i=k$, and $i=l$, respectively, and
\Ref{tshow2}. 

\paragraph{Proof of \Ref{tshow1} for $i\not= k, l$\,:}

The l.h.s.~of \Ref{tshow1} in this case reduces to $\dd_i \ln\Ga /N$
which according to \Ref{XdP} takes the form
\ben
\frac1{N\Ga}\,(\tr[P\dd_i X] + \tr[Z\, \dd_i M]) \;.
\een
Inserting \Ref{MXPZ} we find
\ba
\tr[P\dd_i X] &=& 
\stackrel{1}{\Vvt}{}\!^\iT 
\stackrel{2}{\dd_i\Vv}{}\!^\iT
\stackrel{12}{R}\!'(-\Gl_{kl},\Gl_{kl})\,
\stackrel{1}{\Vw}\,
\stackrel{2}{\Vw}
+
\stackrel{1}{\Vvt}{}\!^\iT 
\stackrel{2}{\Vv}{}\!^\iT
\stackrel{12}{R}\!'(-\Gl_{kl},\Gl_{kl})\,
\stackrel{1}{\Vw}\,
\stackrel{2}{\dd_i\Vw} \;,
\non[2ex]
\tr[Z\dd_i M] &=& 
-\stackrel{1}{\Vvt}{}\!^\iT 
\stackrel{2}{\dd_i\Vv}{}\!^\iT
\stackrel0{\tr}\left[\stackrel{02}{R}{}(-\Gl_{kl},\Gl_{kl})
\stackrel{10}{R}{}\!'\,(0,\Gl_{kl}) \right] 
\stackrel{1}{\Vw}\,
\stackrel{2}{\Vw} 
\non
&&{}
\qquad\qquad
-\stackrel{1}{\Vvt}{}\!^\iT 
\stackrel{2}{\Vv}{}\!^\iT
\stackrel0{\tr}\left[\stackrel{02}{R}{}(-\Gl_{kl},\Gl_{kl})
\stackrel{10}{R}{}\!'\,(0,\Gl_{kl}) \right] 
\stackrel{1}{\Vw}\,
\stackrel{2}{\dd_i\Vw} 
\non
&\stackrel{\Ref{Rrel}}{=}& 
(N\mis 1)\,
\stackrel{1}{\Vvt}{}\!^\iT 
\stackrel{2}{\dd_i\Vv}{}\!^\iT
\stackrel{12}{R}\!'(-\Gl_{kl},\Gl_{kl})\,
\stackrel{1}{\Vw}\,
\stackrel{2}{\Vw}
\non
&&{} 
\qquad\qquad
+
\stackrel{1}{\Vvt}{}\!^\iT 
\stackrel{2}{\Vv}{}\!^\iT
\stackrel{12}{R}\!'(-\Gl_{kl},\Gl_{kl})\,
\Big(N\stackrel{1}{\dd_i\Vw}\,
\stackrel{2}{\Vw}-
\stackrel{1}{\Vw}\,
\stackrel{2}{\dd_i\Vw}\Big) \;.
\nn
\ea
where the crucial role has been played by the relation \Ref{Rrel} of
the quantum $R$-matrix. Together, these expressions combine into
\be
\dd_i\ln\Ga ~=~ \frac1{\Ga}
\left(
\stackrel{1}{\Vvt}{}\!^\iT 
\stackrel{2}{\dd_i\Vv}{}\!^\iT
\stackrel{12}{R}\!'(-\Gl_{kl},\Gl_{kl})\,
\stackrel{1}{\Vw}\,
\stackrel{2}{\Vw}
+
\stackrel{1}{\Vvt}{}\!^\iT 
\stackrel{2}{\Vv}{}\!^\iT
\stackrel{12}{R}\!'(-\Gl_{kl},\Gl_{kl})\,
\stackrel{1}{\dd_i\Vw}\,
\stackrel{2}{\Vw}
\right) \;,
\la{lhs1}
\ee
which gives the l.h.s.~of \Ref{tshow1}. To compute the r.h.s.~of this
equation we note that with \Ref{FFT} the current $\Xi_{\Gl}(\Gl)$ may
be expressed as
\be
\Xi_{\Gl}(\Gl)~:=~ \frac{1}{\alpha(\Gl)} \,
\Big(\TF(\Gl)\, F'(\Gl)- \frac1{N}\,
\tr\!\left[\TF(\Gl)\, F'(\Gl)\right] \Big)
\;.
\la{XIt}
\ee
Since it has its only poles in $\Gl_k$ and $\Gl_l$, it is together
with its twist properties completely determined by its residues
\ba
\stackrel0{\Xi}\!_{\Gl}(\Gl) &=&
\stackrel1{\tr}{}\!\left[
\stackrel{01}{r}{}\!(\Gl\mis\Gl_k)
\;\res{}\!_{\Gl_k}\! 
\Big(\stackrel1{\Xi}\!_{\Gl}(\Gl)\Big)\, 
\right]
+
\stackrel1{\tr}{}\!\left[
\stackrel{01}{r}{}\!(\Gl\mis\Gl_l)
\;\res{}\!_{\Gl_l}\!
\Big(\stackrel1{\Xi}\!_{\Gl}(\Gl)\Big)\, 
\right] 
\non[1ex]
&=& 
\frac1{\Ga}\,
\stackrel{1}{\Vvt}{}\!^\iT 
\stackrel{2}{\Vv}{}\!^\iT 
\left(
\stackrel{12}{R}\!'(-\Gl_{kl},\Gl_{kl})\,
\stackrel{01}{r}\!(\Gl\mis\Gl_l)
-
\stackrel{02}{r}\!(\Gl\mis\Gl_k)
\stackrel{12}{R}\!'(-\Gl_{kl},\Gl_{kl})\,
\right)
\stackrel{1}{\Vw}\,
\stackrel{2}{\Vw} 
\;.
\nn
\ea
The r.h.s.\ of equation \Ref{tshow1} is hence given by
\ba
\lefteqn{\tr\!\left[ A_i\,\Xi_{\Gl}(\Gl_i)\;\right]~=} \non[1ex]
&=&
\frac{1}{\Ga}\,
\stackrel0{\tr}{}\!\left[
\stackrel{1}{\Vvt}{}\!^\iT 
\stackrel{2}{\Vv}{}\!^\iT
\stackrel0{A_i}
\left(
\stackrel{12}{R}\!'(-\Gl_{kl},\Gl_{kl})\,
\stackrel{01}{r}\!(\Gl_{jl})
-
\stackrel{02}{r}\!(\Gl_{jk})
\stackrel{12}{R}\!'(-\Gl_{kl},\Gl_{kl})\,
\right)
\stackrel{1}{\Vw}\,
\stackrel{2}{\Vw} 
\right]
\non
&=&
\frac{1}{\Ga}
\left(
\stackrel{1}{\Vvt}{}\!^\iT 
\stackrel{2}{\dd_i\Vv}{}\!^\iT
\stackrel{12}{R}\!'(-\Gl_{kl},\Gl_{kl})\,
\stackrel{1}{\Vw}\,
\stackrel{2}{\Vw} 
+
\stackrel{1}{\Vvt}{}\!^\iT 
\stackrel{2}{\Vv}{}\!^\iT
\stackrel{12}{R}\!'(-\Gl_{kl},\Gl_{kl})\,
\stackrel{1}{\dd_i\Vw}\,
\stackrel{2}{\Vw} 
\right)
\;.
\la{rhs1}
\ea
Comparing to \Ref{lhs1} proves \Ref{tshow1} for $i\not=k, l$.

\paragraph{Proof of \Ref{tshow1} for $i= k$\,:}

According to its definition in \Ref{varT}, $\Ga$ is a function with
implicit dependence on $\Gl_k$ via the vectors $v$ and $w$, as well as
an explicit dependence due to the explicit appearance of $\Gl_k$ in
the $R$-matrix in \Ref{F}. Accordingly, the l.h.s.~of
\Ref{tshow1} for $i= k$ has contributions originating from the
implicit and explicit dependence of $\Ga$. The former ones appear with
linear dependence on the residues $A_j$, cf.~\Ref{implicit}. In
complete analogy with the computation leading to \Ref{lhs1},
\Ref{rhs1} above, it may be shown that they coincide with the terms on
the r.h.s.\ of \Ref{tshow1} coming from the residue of $\tr[\Xi_\Gl
A]$ at $\Gl_k$.  The remaining terms, i.e.\ those stemming from the
residue of $\tr[\Xi_\Gl^2]$ and the explicit $\Gl_k$ dependence of
$\Ga$, combine into an $A_j$-independent expression, and determine the
function $\betaN(\Gl_{kl})$. Their derivation is slightly more tedious
and we restrict to sketching the essential steps.

For the explicit $\Gl_k$ dependence of $\Ga$ we find from \Ref{XdP}
and after using \Ref{Rrel}
\ba
\dd_k\,\Ga 
&=&
N \stackrel{1}{\Vvt}{}\!^\iT 
\stackrel{2}{\Vv}{}\!^\iT
\Big(\dd_k \stackrel{12}{R}\!'(-\Gl_{kl},\Gl_{kl}) \Big)\,
\stackrel{1}{\Vw}\,
\stackrel{2}{\Vw} 
\non
&&{}
+
\stackrel{0}{\tr} 
\Big[
\stackrel{1}{\Vvt}{}\!^\iT 
\stackrel{2}{\Vv}{}\!^\iT
\stackrel{10}{R}\!^{[1,1]}(0,\Gl_{kl}) \,
\stackrel{02}{R}\!(-\Gl_{kl},\Gl_{kl}) \,
\stackrel{1}{\Vw}\,
\stackrel{2}{\Vw}
\Big] \;.
\la{varkT1}
\ea
To proceed, we reconsider the derivation of \Ref{XdP} but now
interchanging $\Gl_k$ and $\Gl_l$ and the roles of $F$ and
$\widetilde{F}$. With 
\be
\tilde{\Ga}~:=~\ft{\dd}{\dd\Gl}\,[\det \TF]_{\Gl=\Gl_k} ~
\stackrel{\Ref{detF}}{=}~ 
\frac{\Gth_{[00]}(\Gl_{kl})^{N-2}\;\Gth_{[00]}(\ft1N \Gl_{kl})^{N}}
{\Gth_{[00]}((1\!-\!\ft1N) \Gl_{kl})^{N}} \,\Ga^{N-2}  ~\equiv~
h(\Gl_{kl})^{-1}\,\Ga^{N-2} 
\;,
\la{alphatilde}
\ee
a computation similar to the one leading to \Ref{XdP} shows that
\be
\Gd\Ga ~=~ 
\tr[\widetilde{X}\Gd \widetilde{P}] + 
\tr[\widetilde{M}\, \Gd \widetilde{Z}]
-\Ga\,h^{-1}\,\Gd h \;,
\ee
where $\widetilde{M}=\TF(\Gl_k)$, $\widetilde{P}=F(\Gl_k)$, etc., now
correspond to the terms in the expansion analogous to \Ref{MXPZ}
around $\Gl=\Gl_k$. This gives another expression for the explicit
$\Gl_k$-dependence of $\Ga$
\ba
\dd_k\,\tr[PX] &=&
-\stackrel{0}{\tr} 
\Big[
\stackrel{1}{\Vvt}{}\!^\iT 
\stackrel{2}{\Vv}{}\!^\iT
\stackrel{10}{R}\!^{[1,1]}(0,\Gl_{kl}) \,
\stackrel{02}{R}\!(-\Gl_{kl},\Gl_{kl}) \,
\stackrel{1}{\Vw}\,
\stackrel{2}{\Vw}
\Big] -\Ga\,h^{-1}\,\dd_k h \;.
\ea
Combining this with \Ref{varkT1} yields
\ba
\frac1N\,\dd_k \ln \Ga &=& \frac{1}{2\Ga}\, \stackrel{1}{\Vvt}{}\!^\iT 
\stackrel{2}{\Vv}{}\!^\iT
\Big(\dd_k \stackrel{12}{R}\!'(-\Gl_{kl},\Gl_{kl}) \Big)\,
\stackrel{1}{\Vw}\,
\stackrel{2}{\Vw}  - \; \frac1{2N}\,\dd_k \ln h 
\non
&\stackrel{\Ref{g}}{=}&
-\frac{1}{2\Ga}\, \stackrel{1}{\Vvt}{}\!^\iT 
\stackrel{2}{\Vv}{}\!^\iT
\stackrel{12}{R}\!''(-\Gl_{kl},\Gl_{kl}) \,
\stackrel{1}{\Vw}\,
\stackrel{2}{\Vw}  - \; \frac1{2N}\,\dd_k \ln h + \frac{g}{2}\;.
\la{P0}
\ea
We shall compare this expression to the residue of
$\tr\!\left[\Xi_{\Gl}^2(\Gl)\right]$ at $\Gl_k$ from the r.h.s.\ of
\Ref{tshow1}. Making use of the relation \Ref{rRp} and the local
expansion of \Ref{FFT}, the latter reduces to
\be
-\frac1{\Ga}\,
\stackrel{1}{\Vvt}{}\!^\iT 
\stackrel{2}{\Vv}{}\!^\iT
\stackrel{12}{R}\!''(-\Gl_{kl},\Gl_{kl})\,
\stackrel{1}{\Vw}\,
\stackrel{2}{\Vw}  ~-~ 2 f(\Gl_{kl}) + 
\frac{\aaa''(\Gl_l)}{\aaa'(\Gl_l)}
\;.
\la{P1}
\ee
To compute the last term on the r.h.s.\ we recall that the function
$a(\Gl)$ is determined up to a constant by its zero and pole structure
\ben
\aaa(\Gl) ~=~ -\Ga\;
\frac{\Gth_{[00]}(\ft1N \Gl_{kl})\,
\Gth_{[00]}(\ft{N-1}N\Gl_{kl})\,
\Gth_{[00]}(\Gl\!-\!\Gl_k)\,
\Gth_{[00]}(\Gl\!-\!\Gl_l)}
{\Gth_{[00]}(\Gl_{kl})\,
\Gth_{[00]}'(0)\,
\Gth_{[00]}(\Gl\!-\!\Gl_k\!+\!\ft1N \Gl_{kl})\,
\Gth_{[00]}(\Gl\!-\!\Gl_l\!-\!\ft1N \Gl_{kl})}\;,
\een
which in particular gives 
\be
\frac{\aaa''(\Gl_l)}{\aaa'(\Gl_l)} ~=~
2\,\dd_k\ln
\left( \frac{\Gth_{[00]}((1\!-\!\ft1N) \Gl_{kl})^{N/(N-1)}\;
\Gth_{[00]}(\ft1N \Gl_{kl})^N}{  \Gth_{[00]}(\Gl_{kl}) } \right) 
\;.
\la{alphapp}
\ee
Altogether, \Ref{P0} and \Ref{P1} give the differential equation for
the function $\betaN$:
\be
\ln'\betaN ~=~ 
\frac1{2N}\ln'h -\frac{g}{2} - f  + 
\frac{\aaa''(\Gl_l)}{2\aaa'(\Gl_l)} \;.
\ee
Upon inserting the explicit expressions from \Ref{alphatilde},
\Ref{alphapp}, \Ref{rRp}, and \Ref{g}, one verifies that this
equation is precisely solved by $\betaN$ as given in \Ref{beta}
above.

\paragraph{Proof of \Ref{tshow1} for $i= l$\,:} This may be shown in
complete analogy to the computation for $i= k$ above. However, it is
easier to notice that $\sum\,H_j=\sum\,\HH_j=0$ , such that the
$\Gt$-quotient \Ref{varT} can contain no explicit function of $\Gl_l$
alone.

\paragraph{Proof of \Ref{tshow2}\,: }

This is most conveniently shown by starting from the formula derived
in \cite{KoMaSa00}
\ba
\HH_\mu-H_\mu &=& 
\sum_{j=1}^P \res_{\Gl_j} \,
\tr\!\left[ \Xi_{\mu}(\Gl) A(\Gl) \right] 
+\frac12\sum_{j=k,l}\res_{\Gl_j} \,
\tr\!\left[ \Xi_{\mu}(\Gl)\,\Xi_{\Gl}(\Gl)\right]\;,
\la{mulambda}
\ea
with $\Xi_{\Gl}(\Gl)$ from \Ref{XI} and
\ben
\Xi_{\mu}(\Gl) ~:=~
\frac1{\alpha(\Gl)}\,
\Big(\TF(\Gl)\, \frac{d}{d\mu}\, F(\Gl)-
\frac1{N}\,
\tr\!\Big[\TF(\Gl)\, \frac{d}{d\mu}\, F(\Gl)\Big] \Big) \;.
\een
Computations similar to those presented above show that the terms
containing the residues $A_j$ indeed cancel against the corresponding
terms in $\dd_\mu\Ga$ upon using \Ref{SchlesT}, whereas the terms from
the second sum in \Ref{mulambda} precisely match the $\dd_\mu$
variation of the explicit function $\betaN$ given in \Ref{beta}.

This finishes the proof of Theorem~3.2. Let us emphasize once more
that apart from the local arguments which apply independently of the
specific form of the $R$-matrix, the crucial ingredient was the
relation \Ref{Rrel} for $R$.
\qed

\section{One-point Schlesinger transformations}

In the previous section we have constructed elementary two-point
Schlesinger transformations which shift eigenvalues of the residues in
two poles. Arbitrary Schlesinger transformations may be constructed by
successively applying these elementary transformations. In particular,
this allows to construct the elementary one-point transformations
which act in one point $\Gl_k$ only, by combining two transformations
of type \Ref{F} according to
\be
F_{k,k}^{p,q}(\Gl) ~\equiv~
F_{l,k}^{r,q}(\Gl) \,\cdot\, F_{k,l}^{p,r}(\Gl) \;,
\la{1p2}
\ee
each acting in $\Gl_k$ and an auxiliary point $\Gl_l$. The parameters
$v$, $w$ \Ref{vw} of the second transformation now are those found in
the local expansion of the $\Psi$-function transformed under the first
transformation. The dependence of the total $\Gt$-quotient of
\Ref{1p2} on the parameters of the original $\Psi$-function hence is
rather implicit. However, recalling that for a two-point
transformation \Ref{F} the difference of the two poles $\Gl_k$ and
$\Gl_l$ enters as quantum deformation parameter of the $R$-matrix, it
comes as no surprise that the one point transformation \Ref{1p2} also
allows a direct construction in terms of the classical elliptic
$r$-matrix \Ref{r}. This is what we are going to describe in this
section.

Specifically, we consider Schlesinger transformations which act in one
point $\Gl_k$ only, where they lower the eigenvalue of the residue at
in column~$\Ra$ and raise the eigenvalue in column~$\Rb$, i.e.\ act as
\be
T_k ~\ra~ \HT_k ~=~ T_k - P_\Ra  + P_\Rb \;, \qquad
\Ra\not=\Rb \;.
\la{TtrafA}
\ee
\begin{theorem}
The Schlesinger transformation of the system \Ref{SchlesT} which shift
the eigenvalues of the residues $A_j$ according to \Ref{TtrafA} are
given by
\be
\Psi(\Gl)~\ra~ \frac{F(\Gl)}{\sqrt[N]{\det F(\Gl)}}\,\Psi(\Gl) \;,
\la{STA}
\ee
where the $GL(N,\bbC)$-valued matrix $F(\Gl)$ is defined via the
classical elliptic $r$-matrix
\be
\stackrel{0}{F}{}\!\!\stackrel{}{(\Gl)} ~:=~ 
\stackrel{0}{F}{}\!\!_{k,k}^{p,q}\stackrel{}{(\Gl)} ~:=~ 
- y\stackrel{0}{I} ~+~ 
\stackrel1{\Vv}{}\!^\iT \stackrel{01}{r}{}\!\!
\stackrel{}{(\Gl\mis\Gl_k)}
\,\stackrel1{\Vw} \;,
\la{FA}
\ee
where the dependence of $F(\Gl)$ on the parameters of the
$\Psi$-function is completely contained in the vectors $\Vv$, $\Vw$
and the scalar $y$ which are defined as functions of the local
expansion \Ref{asymp} in $\Gl_k$
\be
(\Vv^\iT)^m ~:=~ (G_k^{-1} )_\Ra{}^m \;,\qquad
\Vw_m ~:=~ (G_k)_m{}^\Rb \;,\qquad
y:=~ (Y_k(0))_\Ra{}^\Rb
\;.
\la{vwy}
\ee

\end{theorem}

\paragraph{Proof:} 
As in the proof of Theorem~3.1.\ we need to check twist behavior of
$F(\Gl)$, its local behavior around $\Gl_k$ and the absence of further
singularities induced by \Ref{STA}. The correct twist behavior again
follows from \Ref{Rtwist} since $r$ is obtained in the classical limit
of $R$. The local behavior of \Ref{FA} around $\Gl_k$ is given by
\ba
F(\Gl)&=&\frac1{\Gl\mis\Gl_k}\,\TP + \TZ + (\Gl\mis\Gl_k)\,\TZ_1 
+ \dots 
\non
&\equiv&
\frac1{\Gl\mis\Gl_k}\,
\stackrel1{\Vv}{}\!^\iT 
\stackrel{01}{\Pi}
\;\stackrel1{\Vw}
{}~+~ \Big(
\stackrel1{\Vv}{}\!^\iT 
\stackrel{01}{r_0}
\,\stackrel1{\Vw} ~-~ y\stackrel{0}{I} \Big)  
~+~ (\Gl\mis\Gl_k)\,
\stackrel1{\Vv}{}\!^\iT 
\stackrel{01}{r_1}
\;\stackrel1{\Vw} + \dots  \;.
\la{PZZ}
\ea
Right antisymmetry \Ref{rsym} of $r_0$ together with \Ref{vwy} yields
\be
\TP\,G_k ~=~ \TP\,G_k\, P_\Ra \;,
\quad
\TZ \,G_k \,P_\Rb ~=~ - \TP\,G_k\, P_\Ra\,Y_k(0)\,P_\Rb \;.
\ee
The local expansion \Ref{asymp} then shows that multiplication of
$\Psi$ by $F$ indeed induces the shift \Ref{TtrafA} of $T_k$. Finally,
we note that $\det F(\Gl)$ is a single valued function on the torus
without poles, since $\TP=\res|_{\Gl_k}F(\Gl)$ is nilpotent,
$\TP^2=0$. It is hence a constant, such that the transformation
\Ref{STA} indeed does not induce any additional singularities in
$\Psi$.
\qed

Similar to Theorem~3.2, the change of the $\Gt$-function under
\Ref{STA} may explicitly be integrated in terms of the parameters of
the $\Psi$ function:
\begin{theorem}
Under the transformation \Ref{STA}, the $\Gt$-function changes by a
factor
\be
\frac{\widehat{\tau}}{\tau}  ~=~ \sqrt[N]{\Ga}  ~~\equiv~
\sqrt[N]{\det F} \;,
\la{varTA}
\ee
with the multiplier matrix $F$ from \Ref{FA}.
\end{theorem}

\paragraph{Proof:} 

Recall first that according to the discussion above, $\sqrt[N]{\Ga}
\equiv \sqrt[N]{\det F}$ is indeed a $\Gl$-independent constant. In
analogy to \Ref{tshow1}, \Ref{tshow2}, it remains to show that
\ba
\ft1{N}\, \dd_i \ln \Ga 
&=& \res_{\Gl_i}
\Big(\,\tr\!\left[\, \Xi_{\Gl}(\Gl) A(\Gl)\;\right]+ 
\ft12\,\tr\!\left[\, \Xi_{\Gl}^2(\Gl)\,\right]\,\Big) \;,
\la{tshow1A}
\\[1ex]
\ft1{N}\, \dd_\mu \ln \Ga
&=& -\ft1{2\pi i}\,\oint_a d\Gl
\Big(\,2\, \tr\!\left[\, \Xi_{\Gl}(\Gl) A(\Gl)\;\right]+ 
\tr\!\left[\, \Xi_{\Gl}^2(\Gl)\,\right]\,\Big)
\;,
\la{tshow2A}
\ea
with $\Xi_\Gl$ from \Ref{XI}. Again, we find it convenient to first
construct the inverse Schlesinger transformation $\TF(\Gl) \equiv
F_{k,k}^{q,p}$. According to Theorem~4.1 it is of the form
\be
\stackrel{0}{\TF}{}\!\!\!\stackrel{}{(\Gl)} ~:=~ 
- \tilde{y}\stackrel{0}{I} ~+~ 
\stackrel1{\Vvt}{}\!^\iT \stackrel{01}{r}{}\!\!
\stackrel{}{(\Gl\mis\Gl_k)}
\,\stackrel1{\Vwt} \;,
\la{FTA}
\ee
where the vectors $\Vwt$, $\Vvt$ and the scalar $\tilde{y}$ now depend
on the transformed parameters of the local expansions
\Ref{asymp}. It has a local expansion similar to \Ref{PZZ}
\ba
\TF(\Gl)&=&\frac1{\Gl\mis\Gl_k}\,P + Z + (\Gl\mis\Gl_k)\,Z_1 
+ \dots 
\non
&\equiv&
\frac1{\Gl\mis\Gl_k}\,
\stackrel1{\Vvt}{}\!^\iT 
\stackrel{01}{\Pi}
\;\stackrel1{\Vwt}
{}~+~ \Big(
\stackrel1{\Vvt}{}\!^\iT 
\stackrel{01}{r_0}
\,\stackrel1{\Vwt} ~-~ \tilde{y}\stackrel{0}{I} \Big)  
~+~ (\Gl\mis\Gl_k)\,
\stackrel1{\Vvt}{}\!^\iT 
\stackrel{01}{r_1}
\;\stackrel1{\Vwt} + \dots  \;.
\la{PZZT}
\ea
Moreover, we normalize $\TF(\Gl)$ such that $\det \TF(\Gl)=1$. In
other words, $\TF(\Gl)$ is the matrix of minors of $F(\Gl)$
\be
F(\Gl)\,\TF(\Gl)~=~ \Ga\,I\, \;,
\la{FFTA}
\ee
and
\be
\Ga~=~\det F(\Gl) ~=~ y\tilde{y} ~+~ 
\stackrel1{\Vvt}{}\!^\iT \stackrel2{\Vv}{}\!^\iT 
\stackrel{12}{r_1}{}\!\!
\,\stackrel1{\Vwt}\stackrel2{\Vw} \;.
\la{alphaA}
\ee
{}From the local expansion of \Ref{FFTA} it follows that as for the
two-point transformations $\Vwt\sim\Vw$ and may hence be normalized to
$\Vwt=\Vw$. The isomonodromic dependence of the vectors
$\Vv$, $\Vw$ again follows from \Ref{implicit}. In addition,
\Ref{Alocal} together with \Ref{SchlesT} implies that 
\ba
\dd_j y  &=& 
-\Vv^\scT \; \sum_{(\iA,\iB)\,\not=(0,0)}
A_j^{\iA\iB}\,\Gs_{\iA\iB}\,w'_{\iA\iB}(\Gl_{kj}) \; \Vw \;,
\qquad j\not=k\;,
\non
\dd_k y  &=& 
\Vv^\scT \; \sum_{j\not=k}\sum_{(\iA,\iB)\,\not=(0,0)} 
A_j^{\iA\iB}\,\Gs_{\iA\iB}\,w'_{\iA\iB}(\Gl_{kj}) \; \Vw \;.
\ea
We can now compute the
l.h.s.\ of \Ref{tshow1A} for $i\not=k$:
\ba
\frac1{N}\,
\dd_i \ln \Ga &=& \frac1{\Ga}\, \tr \left[ \TF(\Gl)\,\dd_i F(\Gl)
\right]
\non
&=& \frac1{N\Ga} 
\Big(
\tr [P\,\dd_i \TZ_1] + \tr [Z\,\dd_i \TZ] + \tr [Z_1\,\dd_i \TP]
\Big)
\non
&=& 
\frac1{N\Ga} 
\Big(
2\stackrel1{\Vvt}{}\!^\iT \!\stackrel2{\dd_i\Vv}{}\!^\iT \!
\stackrel{12}{r_1}{}\!\!
\,\stackrel1{\Vw}\stackrel2{\Vw} 
+
2\stackrel1{\Vvt}{}\!^\iT \!\stackrel2{\Vv}{}\!^\iT \!
\stackrel{12}{r_1}{}\!\!
\,\stackrel1{\Vw}\stackrel2{\dd_i\Vw} 
+ N \, \tilde{y}\, \dd_i y + 
\non
&&\quad 
+ \stackrel{0}{\tr} \Big[
\stackrel1{\Vvt}{}\!^\iT \!\stackrel2{\dd_i\Vv}{}\!^\iT \!
\stackrel{01}{r_0}{}\stackrel{02}{r_0}{}\!\!
\,\stackrel1{\Vw}\stackrel2{\Vw} \Big]
+ \stackrel{0}{\tr} \Big[
\stackrel1{\Vvt}{}\!^\iT \!\stackrel2{\Vv}{}\!^\iT \!
\stackrel{01}{r_0}{}\stackrel{02}{r_0}{}\!\!
\,\stackrel1{\Vw}\;\stackrel2{\dd_i\Vw} \Big]
\;\Big)
\non
&=&
\frac{1}{\Ga}\,
\Big( \tilde{y} \,\dd_i y ~+~
\stackrel{1}{\Vvt}{}\!^\iT \!
\stackrel{2}{\dd_i\Vv}{}\!^\iT
\stackrel{12}{r_1}\;
\stackrel{1}{\Vw}\,
\stackrel{2}{\Vw}  ~+~
\stackrel{1}{\Vvt}{}\!^\iT \!
\stackrel{2}{\Vv}{}\!^\iT
\stackrel{12}{r_1}\;
\stackrel{1}{\dd_i\Vw}\,
\stackrel{2}{\Vw} 
\Big) \;,
\la{lhs2}
\ea
where for the last equation we have made use of the property
\Ref{rrel} of the classical $r$-matrix. To compute the r.h.s.\ of
\Ref{tshow1A} we note that $\Xi_\Gl((\Gl)$  has its only poles (of
first and second order) in $\Gl_k$. Together with its twist properties
it is hence completely determined by its residues
\ba
\stackrel0{\Xi_{\Gl}}(\Gl) &=&
-\frac{\tilde{y}}{\Ga}\,
\stackrel{1}{\Vv}{}\!^\iT 
\stackrel{01}{r'}\!(\Gl\mis\Gl_k)
\stackrel{1}{\Vw} ~-~
\frac1{\Ga}\,
\stackrel{1}{\Vvt}{}\!^\iT 
\stackrel{2}{\Vv}{}\!^\iT
\Big( 
\stackrel{02}{r}\!(\Gl\mis\Gl_k)
\stackrel{12}{r_1}-
\stackrel{12}{r_1}\;
\stackrel{01}{r}\!(\Gl\mis\Gl_k)\,
\Big)
\stackrel{1}{\Vw}\,
\stackrel{2}{\Vw} 
\nn
\ea
The r.h.s.\ of \Ref{tshow1A} hence gives
\ba
\lefteqn{\tr\!\left[ A_i\,\Xi_{\Gl}(\Gl_i)\;\right]~=} \non[1ex]
&=&
- \frac{\tilde{y}}{\Ga}\, 
\stackrel{0}{\tr} \left[
\stackrel{1}{\Vv}{}\!^\iT 
\stackrel{0}{A_i}\,
\stackrel{01}{r'}\!(\Gl_{jk})
\stackrel{1}{\Vw} \right]
+ \frac1{\Ga}\,
\Big(
\stackrel{1}{\Vvt}{}\!^\iT \!
\stackrel{2}{\dd_i\Vv}{}\!^\iT
\stackrel{12}{r_1}\;
\stackrel{1}{\Vw}\,
\stackrel{2}{\Vw} 
+
\stackrel{1}{\Vvt}{}\!^\iT \!
\stackrel{2}{\Vv}{}\!^\iT
\stackrel{12}{r_1}\;
\stackrel{1}{\dd_i\Vw}\,
\stackrel{2}{\Vw} 
\Big)
\non
&=&
\frac1{\Ga}\,
\Big( \tilde{y} \,\dd_i y ~+~
\stackrel{1}{\Vvt}{}\!^\iT \!
\stackrel{2}{\dd_i\Vv}{}\!^\iT
\stackrel{12}{r_1}\;
\stackrel{1}{\Vw}\,
\stackrel{2}{\Vw}  ~+~
\stackrel{1}{\Vvt}{}\!^\iT \!
\stackrel{2}{\Vv}{}\!^\iT
\stackrel{12}{r_1}\;
\stackrel{1}{\dd_i\Vw}\,
\stackrel{2}{\Vw} 
\Big) \;.
\la{rhs2}
\ea
Comparing this to \Ref{lhs2} proves \Ref{tshow1A} for $i\not=k$. The
$\Gt$-quotient \Ref{varTA} is hence determined up to an explicit
function of $\Gl_k$ and the modulus of the torus $\mu$. Again, $\sum
H_j = \sum \HH_j = 0$ rules out a possible explicit function of
$\Gl_k$. This proves \Ref{tshow1A} for $i=k$. Finally, equation
\Ref{tshow2A}, i.e. absence of an explicit function of the modulus
$\mu$ in \Ref{varTA}, is shown by starting from \Ref{mulambda} and
similar computations to those presented above.
\qed

\section*{Acknowledgements:}

We wish to thank D.~Korotkin and P.~Kulish for numerous enlightening
discussions and helpful comments on the manuscript. This work has been
supported in part by the Portuguese Foundation for Science and
Technology under POCTI/33858/MAT/2000.

\begin{appendix}

\mathon
\section{Appendix: Elliptic functions and $R$-matrix}
\mathoff

The building blocks of the elliptic $R$-matrix are the
$\mathfrak{sl}_{N,\bbC}$-valued matrices
\be
\Gs_{\iA\iB}~:=~ h^\iA g^\iB\;,\qquad 
\ee 
where $g$ and $h$ are the cyclic matrices 
\be 
g_m{}^n=\Gd_m^n\,\Gw^{m-1}\;, \quad h_m{}^n=
\Gd_{m+1}^n\;,\quad \Go=e^{2\pi\I / N} \;, 
\la{gh}
\ee 
satisfying the relations $\Gw gh= hg$\,, and $g^N = h^N = I$\,. We
further define
\be
\Gs^{\iA\iB}:=\frac{\Gw^{-\iA\iB}}{N}\,\Gs_{-\iA,-\iB}\;, \qquad
\mbox{such that} \;\;\;
\tr\left[\,\Gs_{\iA\iB}\,\Gs^{\iC\iD}\,\right]=
\Gd_{\iA}^{\iC}\,\Gd_{\iB}^{\iD}\;.  
\la{sigo}
\ee
Define the elliptic functions
\be
\Gth_{[\iA\iB]}(\Gl;\mu)~:=~
\GtH\left[\ft{\iA}{N}\mis\ft12\,,\ft12\mis\ft{\iB}{N}\right](\Gl;\mu)\;,
\ee
where
\be
\GtH\left[p,q\right](\Gl;\mu):=
\sum_{m\in\bbZ} e^{ \I\pi (m\pls p)^2\mu+2\I\pi(m+p)(\Gl+q) }
\;,
\ee
are the usual Jacobi theta functions with rational
characteristics. Define further
\be
W_{\iA\iB}(\Gl,\zeta;\mu):= 
\frac{\Gth_{[\iA\iB]}(\Gl+\ft{\zeta}{N};\mu)}
{\Gth_{[\iA\iB]}(\ft{\zeta}{N};\mu)} \;,
\la{W}
\ee
and
\be
w_{\iA\iB}(\Gl;\mu):= 
\frac{\Gth_{[\iA\iB]}(\Gl;\mu)\,\Gth_{[00]}'(0;\mu)}
{\Gth_{[\iA\iB]}(0;\mu)\,\Gth_{[00]}(\Gl;\mu)} \;,\qquad
\mbox{for}\quad (\indA,\indB)\not=(0,0) \;,
\la{w}
\ee
such that the $w_{\iA\iB}$ have a simple pole with unit residue at
$\Gl\equ0$ and twist properties
\ba
w_{\iA\iB}(\Gl\pls1;\mu) \,\Gs_{\iA\iB} &=&
-\Gw^\iA\,w_{\iA\iB}(\Gl;\mu) \,\Gs_{\iA\iB}~=~
w_{\iA\iB}(\Gl;\mu)\;g^{-1}\,\Gs_{\iA\iB}\,g
\;,
\non
w_{\iA\iB}(\Gl\pls\mu;\mu) \,\Gs_{\iA\iB} &=&
-\Gw^\iB\,w_{\iA\iB}(\Gl;\mu) \,\Gs_{\iA\iB}~=~
w_{\iA\iB}(\Gl;\mu)\;h^{-1}\,\Gs_{\iA\iB}\,h
\;.
\la{bas}
\ea
The combinations $w_{\iA\iB}(\Gl;\mu) \,\Gs_{\iA\iB}$ hence provide a
basis for the connections twisted according to \Ref{twist}. For
simplicity in our notation, we suppress the explicit $\mu$-dependence
of all these functions in the main text. Further define the functions
\be
\CZ_{\iA\iB}(\Gl;\mu):= 
\frac{w_{\iA\iB}(\Gl;\mu)}{2\pi\I}
\left(
\frac{\Gth_{[\iA\iB]}'(\Gl;\mu)}{\Gth_{[\iA\iB]}(\Gl;\mu)} -
\frac{\Gth_{[\iA\iB]}'(0;\mu)}{\Gth_{[\iA\iB]}(0;\mu)} 
\right) \;,
\qquad
(\indA,\indB)\not=(0,0) \;,
\la{Z}
\ee
which have no poles and twist behavior according to
\ben
\CZ_{\iA\iB}(\Gl\pls1;\mu) = -\Gw^\iA\,\CZ_{\iA\iB}(\Gl;\mu)
\;,\quad
\CZ_{\iA\iB}(\Gl\pls\mu;\mu) = -\Gw^\iB\,
\left(\CZ_{\iA\iB}(\Gl;\mu)-w_{\iA\iB}(\Gl;\mu)\right) \;.
\een
In particular, 
\be
\dd_\mu w_{\iA\iB}(\Gl;\mu) ~=~ \dd_\Gl \CZ_{\iA\iB}(\Gl;\mu) \;.
\ee
Belavin's
elliptic $R$-matrix \cite{Bela81} is defined as
\be
\stackrel{12}{R}{}(\Gl,\zeta;\mu)~:=~ 
\frac{\Gth_{[00]}(\frac{\zeta}{N};\mu)}
{\Gth_{[00]}(\Gl\pls\frac{\zeta}{N};\mu)}
\sum_{(\iA,\iB)\,\in\, \bbZ_N\times\bbZ_N}\!\!\! 
W_{\iA\iB}(\Gl,\zeta;\mu)
\stackrel1{\Gs}{}\!\!_{\iA\iB}\stackrel2{\Gs}{}\!^{\iA\iB}
\;,
\la{R}
\ee
with the functions $W_{\iA\iB}$ from \Ref{W} and the usual tensor
notation
\be
\stackrel1{A} \;= 
A\otimes I \;,\quad \stackrel2{A} \;= I\otimes A\;,\quad
\mbox{etc.}
\la{tensor}
\ee
It has twist properties following from \Ref{bas}
\ba
\stackrel{12}{R}{}(\Gl+1,\zeta;\mu) &=&
\stackrel1{g}{}\!^{-1}\stackrel{12}{R}{}(\Gl,\zeta;\mu)\,
\stackrel1{g} ~=~
\stackrel2{g}\;\stackrel{12}{R}{}(\Gl,\zeta;\mu)\,
\stackrel2{g}{}\!^{-1}
\;,
\non 
\stackrel{12}{R}{}(\Gl+\mu,\zeta;\mu) &=&
\stackrel1{h}{}\!^{-1}\stackrel{12}{R}{}(\Gl,\zeta;\mu)\,
\stackrel1{h} ~=~
\stackrel2{h}\;\stackrel{12}{R}{}(\Gl,\zeta;\mu)\,
\stackrel2{h}{}\!^{-1}
\;.
\la{Rtwist}
\ea
Invariance of the $R$-matrix under simultaneous conjugation with $g$
or $h$ in both tensor spaces is referred to as $\bbZ_N$ symmetry. In
components, this implies that
\be
R_{ab}^{cd}{}(\Gl,\zeta;\mu) ~\equiv~ \frac1{N}\,\Gd_{a+b}^{c+d}\, 
S^{a-c,a-d}(\Gl,\zeta;\mu) \;.
\la{RZsym}
\ee
It is useful to note that the components $S^{ab}$ allow a product
representation \cite{RicTra86} according to
\ba
S^{ab}(\Gl,\zeta;\mu)&=&
\chi(\Gl,\zeta;\mu)\;
\frac{\Gth_{[b-a,0]}(\Gl+\zeta;N\mu)}
{\Gth_{[-a,0]}(\zeta;N\mu)\,\Gth_{[b,0]}(\Gl;N\mu)}
\;,
\la{Rprod}\\[1ex]
\chi(\Gl,\zeta;\mu)&:=&
\frac{\Gth_{[00]}(\Gl;\mu)\,\Gth_{[00]}(\frac{\zeta}{N};\mu)}
{\Gth_{[00]}(\Gl\pls\frac{\zeta}{N};\mu)}\;
\frac{\Gth_{[00]}'(0;N\mu)}{\Gth_{[00]}'(0;\mu)}
\;.
\nn
\ea
The elliptic $R$-matrix satisfies further standard properties such as
the Yang-Baxter equation, unitarity, crossing symmetry (see
\cite{Bela81,ChuChu81,Cher82,Bovi83,Trac85,AFRS99}). In the
main text we make use of its standard value at zero
\be
R(0,\zeta;\mu) ~=~ \Pi\;,\qquad
\mbox{with}\quad \Pi_{k\,l}^{mn} ~=~ \Gd_k^n\,\Gd_l^m \;,
\la{R0}
\ee
and the reflection property
\be
\stackrel{12}{R}{}(\Gl,\zeta;\mu) ~=~ 
\stackrel{21}{R}{}(-\Gl,-\zeta;\mu) \;.
\la{Ras}
\ee
The classical limit of $R$ corresponds to sending $\zeta\ra0$ 
\be
N R(\Gl,\zeta;\mu) ~=~ 
I + \zeta\,r(\Gl;\mu) + \CO\left(\zeta^2\right)  \;,
\la{zeta0}
\ee
and yields the classical $r$-matrix
\be
\stackrel{12}{r}{}(\Gl;\mu)~=~ 
\sum_{(\iA,\iB)\,\not=(0,0)}\!\!\! 
w_{\iA\iB}(\Gl;\mu)
\stackrel1{\Gs}{}\!\!_{\iA\iB}\stackrel2{\Gs}{}\!^{\iA\iB} ~=~
\frac1{\Gl}\stackrel{12}{\GO}  +
\stackrel{12}{r}{}\!\!\!_0 + 
\Gl\! \stackrel{12}{r}{}\!\!\!_1~+~ \CO\left(\Gl^2\right)
\;.
\la{r}
\ee
with 
\be
\GO=\Pi - \ft1{N} I \;,\qquad
\stackrel{12}{r}{}\!\!\!_0= \sum_{(\iA,\iB)\,\not=(0,0)}
\frac{\Gth'_{[\iA\iB]}(0;\mu)}{\Gth_{[\iA\iB]}(0;\mu)}\;
\stackrel1{\Gs}{}\!\!_{\iA\iB}\stackrel2{\Gs}{}\!^{\iA\iB}
\;,\qquad
\mbox{etc.}\;,
\ee
and the functions $w_{\iA\iB}(\Gl;\mu)$ defined in \Ref{w} above. To
compute the variation of the $\Gt$-function in the main text we will
need another relation satisfied by the $R$-matrix 
\ba
\stackrel0{\tr}\left[\stackrel{10}{R}{}(-\zeta,\zeta;\mu)
\stackrel{02}{R}{}\!'\,(0,\zeta;\mu) \right] 
&=&
\stackrel{12}{R}{}\!'\,(-\zeta,\zeta;\mu) \;
(\stackrel{12}{I}-\stackrel{}{N}\,\stackrel{12}{\Pi})
\non
&&{} \qquad\qquad +
N \varphi(\zeta;\mu)\stackrel{12}{R}{}(-\zeta,\zeta;\mu)
\;,
\la{Rrel}
\ea
which may be proven by rather tedious but straight-forward
computation, checking twist behavior, residues and an additive
constant. The function $\varphi(\zeta;\mu)$ on the r.h.s.\ depends on
the specific normalization of the $R$-matrix in
\Ref{R}. It may be expressed in terms of theta functions, however the
explicit form is not of particular interest for this text. The
classical limit of \Ref{Rrel} gives rise to
\ba
\stackrel0{\tr}\Big[\stackrel{01}{r_0}\;
\stackrel{02}{r_0}\Big] 
&=&
\stackrel{12}{r_1} 
(\stackrel{}{N}\,\stackrel{12}{\Pi}-\stackrel{}{2}\stackrel{12}{I})
\;,
\la{rrel}
\ea
for the coefficients $r_0$, $r_1$ of the classical $r$-matrix. This
relation crucially enters the computation of the variation of the
$\Gt$-function under one-point transformations.

{}From the product representation \Ref{Rprod} one finds
\be
R(\Gl,\zeta;\mu) ~=~ 
\frac{\chi(\Gl,\zeta;\mu)}{\chi(\zeta,\Gl;\mu)}\;\Pi\,R(\zeta,\Gl;\mu)
\;.
\la{stra}
\ee
In particular, this gives rise to
\be
R(-\zeta,\zeta;\mu)\,\bbP_+ ~=~ 0 ~=~ 
\bbP_-\, R(\zeta,\zeta;\mu)   \;,
\la{Rsym}
\ee
with the projection operators $\bbP_\pm=\ft12(I\pm\Pi)$. Existence of
a right antisymmetric singular point at $\Gl=-\zeta$ proves to be
essential for the construction of the two-point Schlesinger
transformation. From the classical limit of these equations one
further finds that
\be
r_0\,\bbP_+ ~=~ 0 ~=~ \bbP_-\,r_0 \;,
\la{rsym}
\ee
which will be of similar importance for the one-point
transformations. Equation \Ref{stra} further gives rise to
\be
r(\Gl;\mu) ~=~
\Pi\,R'(0,\Gl;\mu) + f(\Gl)\,I \;,
\la{rRp}
\ee
with the scalar function
\ben
f(\Gl)=\frac{\dd}{\dd \Gl}\, \ln 
\left( \frac{\Gth_{[00]}(\ft1N \Gl) ^N}{\Gth_{[00]}(\Gl) ^{1/N}} \right) 
\;.
\een
Similarly, one obtains
\be
R^{[1,1]}(-\Gl,\Gl) \,\bbP_+ ~=~ g(\Gl)\,R'(-\Gl,\Gl) \,\bbP_+ \;,
\la{g}
\ee
with scalar function
\ben
g(\Gl)~=~
\frac{\dd}{\dd \Gl}\, \ln 
\left( \frac{\Gth_{[00]}((1\!-\!\ft1N) \Gl)^{(N+1)/(N-1)}\;
\Gth_{[00]}(\ft1N \Gl)}{  \Gth_{[00]}(\Gl) } \right) \;.
\een

\end{appendix}

\end{document}